\input{aipcheck}

\newcommand{\captn}[1]{\caption{#1}}
\newcommand{\ga}{\alpha}
\newcommand{\gd}{\delta}
\newcommand{\gf}{\phi}
\newcommand{\gc}{\chi}
\newcommand{\gk}{\kappa}
\newcommand{\gl}{\lambda}
\newcommand{\gm}{{\mu}}
\newcommand{\gth}{\theta}
\newcommand{\gp}{\pi}
\newcommand{\gF}{\Phi}
\newcommand{\gL}{\Lambda}
\newcommand{\gTh}{\Theta}
\newcommand{\cL}{{\cal L}}
\newcommand{\cW}{{\cal W}}
\newcommand{\ua}{{\underline a}}
\newcommand{\ub}{{\underline b}}
\newcommand{\uc}{{\underline c}}
\newcommand{\bff}{{\bar f}}

\newcommand{\Tr}{\mbox{Tr}}

\newcommand{\ra}{\rightarrow}
\newcommand{\inv}{^{-1}}
\newcommand{\dsp}{\displaystyle}
\newcommand{\labl}[1]{\label{#1}}
\newcommand{\Kh}{K\"{a}hler}
\newcommand{\tabu}[2]{\begin{tabular}{#1} #2 \end{tabular}}
\newcommand{\non}{\nonumber}
\newcommand{\bC}{{\overline C}}
\newcommand{\bD}{{\overline D}}
\newcommand{\bI}{{\overline I}}
\newcommand{\bJ}{{\overline J}}
\newcommand{\bW}{{\overline W}}
\newcommand{\bgf}{{\bar\phi}}
\newcommand{\bgc}{{\bar\chi}}
\newcommand{\bgm}{{\bar\mu}}
\newcommand{\bgth}{{\bar\theta}}
\newcommand{\bgF}{{\overline\Phi}}
\newcommand{\bgTh}{{\overline\Theta}}
\newcommand{\MSbar}{$\overline{\mbox{MS}}$}

\newcommand{\sfrac}[2]{\mbox{$\frac{#1}{#2}$}}

\documentclass[
    ,final            
  ]
  {aipproc}

\layoutstyle{6x9}

\usepackage{verbatim}
\usepackage{amsfonts}
\usepackage{latexsym}
\usepackage{bbm}
\usepackage{graphics}

\begin{document}
\begin{flushright}
IC/2006/085\\
HD-THEP-06-22\\
SIAS-CMTP-06-6
\end{flushright}
\title{Two Loop Effective K\"ahler Potential}


\author{Tino S. Nyawelo}{
  address={
The Abdus Salam ICTP,
Strada Costiera 11, I--34014 Trieste, Italy }
}

\author{Stefan Groot Nibbelink }{
  address={ 
Institut f\"ur Theoretische Physik, Universit\"at Heidelberg,
Philosophenweg 16 und 19, D-69120 Heidelberg, Germany
}
,altaddress={
Shanghai Institute for Advanced Study, 
University of Science and Technology of China,  
99~Xiupu~Rd, Shanghai 201315, P.R.~China}
}

\begin{abstract}
In this talk we study the renormalization of the effective K\"ahler
potential at one and two loops for general four dimensional
(non--renormalizable) ${\cal{N}}=1$ supersymmetric theories described
by arbitrary \Kh\ potential, superpotential and gauge kinetic function.  
We consider the Wess-Zumino model as an example. 
 \end{abstract}

\maketitle


\section{Introduction}
\label{sc:intro}

The use of the effective action in quantum field theories has proven
to be a powerful tool. The effective action of a supersymmetric theory
is encoded by three functions of the chiral multiplets. The
superpotential and the gauge kinetic function are holomorphic and
consequently much constraint. This is reflected in various 
non--renormalization theorems \cite{Grisaru:1979wc} for 
these functions and lead to results to all order
\cite{Novikov:1982px,Shifman:1986zi}, and even non--perturbative
results \cite{Seiberg:1993vc,Intriligator:1994jr}.  The third
function, the real K\"ahler potential receives corrections at all
orders in perturbation theory. Precisely, because for the K\"ahler
potential there are no non--renormalization theorems, it is interesting to
understand its higher order quantum corrections. The computation
of the effective K\"ahler potential can be used in phenomenological
applications, as it encodes the wave function renormalization of
chiral multiplets. The physical masses can only be determined when the
effect of wave function renormalization is taken into account. A
direct computation of the wave function renormalization quickly
corresponds to an order of a hundred (super)graphs, while even in
the most general case we needed to compute only about a dozen.

In this talk we review supergraph computation of the \Kh\ 
potential up to two loops including gauge interaction \cite{Nibbelink:2005wc}. 
The calculation was performed in a background of chiral multiplets. To
avoid mixing between the quantum chiral and vector multiplets a
supersymmetric variant of the Feynman--'t Hooft gauge fixing was
employed. The divergent one and two loop supergraphs were regularized
using the $\overline{\rm DR}$ scheme. Before, only some partial
results have been obtained at two loop, for the renormalizable
Wess--Zumino model with a cubic superpotential \cite{Buchbinder:1994xq}. 
The extension to a general non--renormalizable Wess--Zumino
\cite{Buchbinder:2000ve} did not look covariant and formed part of the
motivation for our work.  Because the two loop result is rather
complicated, we use the Wess--Zumino model as an illustration.

\section{Gauged non-linear supersymmetric sigma model}

We start with a general $N=1$ supersymmetric non--linear sigma model of
(anti--)chiral superfields $\gf^a(\bgf_\ua)$ described by a \Kh\ 
potential $K(\bgf,\gf)$ and a superpotential $W(\gf)$. 
$G^\ua{}_a ~=~ K_,{}^\ua{}_a$ is the resulting \Kh\ metric. 
Some of the linear isometries are gauged using vector superfields  $V
= V^I T_I$. The algebra of these Hermitean generators $T_I$ is encoded
in purely imaginary structure coefficients $c^K{}_{IJ}$. 
The classical Lagrangian is given by: 
\begin{equation}
\cL_{cl} ~=~\frac 12\, \int\mathrm{d}^4\gth\, 
K(\bgf,e^{2V}\gf) 
\,+\, 
\int \mathrm{d}^2\gth\,\Bigl(W(\gf)
+ f_{IJ}(\gf)\cW^{I\,\ga}\cW^{J}{}_\ga\Bigl)~+~\rm{h.c.} 
\labl{clKh} 
\end{equation}
The gauge kinetic function $f_{IJ}$ is symmetric in the gauge indices
$I$ and $J$. Its real part $h_{IJ} = \frac 12 ( f_{IJ} + \bar f_{IJ} )$ 
is equal to the gauge coupling $1/g^2$. The super gauge field strengths are 
given by $\cW_\ga = -\frac 18\,{\bD{}^2}(e^{-2V}D_\ga e^{2V})$.

The quantum corrections to the \Kh\ potential are determined by using
background field method: One expans the theory around some 
non--trivial background $\gf$ for the chiral multiplets by replacing
$\gf ~\ra~ \gf + \gF$, while  assuming a trivial background for the gauge sector. 
Only the quantum fields $\gF$ and $V$ appear in the loops. Gauge
fixing is performed by introducing the Faddeev--Poppov (FP) ghosts
$C,C'$, and Nielsen--Kallosh (NK) ghost $\gc$. The gauge fixing Lagrangian 
\begin{equation}
\cL_{GF} ~=~ - 
\int \mathrm{d}^4\gth\, h_{IJ}\,\bgTh{}^I \gTh^J~,\,~
\gTh^I ~=~ -\frac {\sqrt 2}4 \bD^2\Big( 
V^I  \,+\, (h\inv)^{IJ} K^{\ua}{}_a (T_J \gf)^a \, \frac{1}{\Box} \bgF_\ua
\Big) 
\labl{GFac}
\end{equation}
is a generalization of the Feynman--'t~Hooft gauge fixing for
spontaneously broken gauge theories \cite{Ovrut:1981wa}.  
The combined ghost Lagrangian reads 
\begin{equation}
\cL_{FP + NK} ~=~ - \int \mathrm{d}^4\gth\, h_{IJ}\,\bgc{}^I \gc^J~ +
\frac 1{\sqrt 2}\int\mathrm{d}^2\gth \, C_I' \gd_\gL \gTh^I + \frac
1{\sqrt 2}\int\mathrm{d}^2\bgth \, \bC_I' \gd_\gL \bgTh^I~, 
\end{equation}
where $\gd_\gL \gTh^I$ denotes the variation of the gauge.

\begin{figure}
\raisebox{0ex}{\scalebox{0.45}{\mbox{\input{bubbles.pstex_t}}}}
\captn{
Summing all bubble diagrams give the  one loop
effective \Kh\ potential. 
}
\labl{fg:1Lbubble}
\end{figure}

\section{One and two loops effective K\"ahler potential}

The one loop calculation of the effective \Kh\ potential involves the
computation of one loop vacuum bubble graphs with multiple insertions
of two--point interactions, as indicated in figure \ref{fg:1Lbubble}. 
The effective one loop \Kh\ potential is given by  
\begin{equation}
K_{1L} ~=~ 
- \frac 1{16\gp^2} \, 
\Tr\, h\inv M_C^2 \Big(2 - \ln \frac {h\inv M_C^2}{\bgm^2} \Big)
~+~ \frac 1{32\gp^2}\, 
\mathrm{tr}\, M_W^2 \Big(2 - \ln \frac {M_W^2}{\bgm^2} \Big)~, 
\labl{1LKh}
\end{equation}
where $\bgm^2 ~=~ 4\gp e^{-\gamma} \gm^2$ defines the \MSbar\
renormalization scale, $(M_C^2)_{IJ} = 2\,\bgf T_I G T_J \gf$ 
is mass matrix of the FP ghosts, and finally chiral superfield mass
matrix $M_W^2 = G\inv \bW (G\inv)^T W$  results from 
the superpotential.  The traces over the chiral multiplet and the
adjoint representation are denoted by  $\mathrm{tr}$ and $\Tr$,
respectively. Our one loop results are consistent with the existing
literature \cite{Grisaru:1996ve,Pickering:1996gt,Buchbinder:1999jw, 
Brignole:2000kg}.\footnote{In the non--Abelian case the ghost mass 
$M_C^2$ and vector mass $M_V^2 = \frac 12( M_C^2 + {M_C^2}^T)$ are not
equal anymore, so that our results slightly deviate from these
reference which use a different gauge.}

\begin{figure}
\tabu{ccc}{
\raisebox{.5ex}{\scalebox{0.5}{\mbox{\input{sc8graph.pstex_t}}}}
&\qquad \qquad &
\raisebox{0ex}{\scalebox{0.3}{\mbox{\input{sc0-graph.pstex_t}}}}
\\[2ex]
a&& b 
}
\captn{The two loop diagrams either have the  ``8'' (figure a)
or ``$\ominus$'' (figure b) topology.   
}
\labl{fg:2Lgraphs}
\end{figure}

At the two loop level there are two different topologies of the
supergraphs that may contribute to the \Kh\ potential, see in figure
\ref{fg:2Lgraphs}. These two loop supergraphs are reduced to two loop
scalar graphs using standard supergraph techniques. Being two loop
graphs, these integrals contain subdivergences. The subtraction of
subdivergences is crucial because otherwise one would not obtain local
counter terms from the two loop level onwards. The expressions of
two loops scalar integral ``8'' (figure \ref{fg:2Lgraphs}.a) and
``$\ominus$''  (figure \ref{fg:2Lgraphs}.b) with their subdivergences
and poles subtracted off can be computed using the methods described in 
\cite{Ford:1992pn}, and are denoted respectively by $\bJ$ and $\bI$
\cite{Nibbelink:2005wc}. The full two \Kh\ potential is given by 
\begin{eqnarray}
K_{2L} &=& 
\frac 12 \, R^\ua{}_a{}^\ub{}_b ~ \overline{J}{}^a{}_\ua{}^b{}_\ub(M^2,M^2) 
+ 
\frac 16\, \bW^{;\ua\,\ub\,\uc} \,W_{;abc} ~ 
\overline{I}{}^a{}_\ua{}^b{}_\ub{}^c{}_\uc(M^2,M^2,M^2)  
\non \\[0ex] 
&&
+ \frac 12 \, h_{LP}\,c^P{}_{IN}\, h_{JQ}\,c^Q{}_{KM}\, 
\Big\{ 
 \bI{}^{IJKLMN}({M_C^2}, M_C^2, M_V^2) 
- \overline{I}{}^{IJKLMN}(M_C^2, {M_C^2}^T, M_V^2)
\Big\} 
\non \\[0ex] 
& &
- (G T_I \gf)^\ua{}_{;a} \, (\bgf T_J G)_b{}^{;\ub}\, 
\overline{I}{}^a{}_\ua{}^b{}_\ub{}^{IJ}(M^2,M^2,M_V^2)
\non \\[0ex] 
& &
+ \frac 18 \, 
f_{IK\, a} \, \bff_{JL\,}{}^{\ua}
\Big\{
2\, h\inv{}^{KL}\, \overline{J}{}^a{}_{\ua}{}^{IJ}(M^2, M_V^2) 
- G\inv{}^a{}_\ua\, \overline{J}{}^{IJKL}(M_V^2, M_V^2) 
\non \\[0ex] 
& &
+ (T_M \gf)^a(\bgf T_N)_\ua\overline{I}{}^{IJKLMN}(M_V^2,M_V^2,M_C^2)
\Big\} 
+ \frac 18
\Big\{ 
f_{IKb}(G\inv\bW)^{b\ua}\bff_{JL}{}^{\ub}({G\inv}^T W)_{\ub a} 
\non \\[0ex]
& &
- f_{MK\, a} \, \bff_{NL\,}{}^{\ua}
\Big(
\gd^M{}_I (h\inv M_V^2)^N{}_J  
~+~ 
\gd^N{}_J (h\inv M_V^2)^M{}_I
\Big) 
\Big\} \, \overline{I}{}^a{}_{\ua}{}^{IJKL}(M^2,M_V^2,M_V^2)
\non \\[0ex]
& &
+ \frac 12 \Big( 
f_{IK\, a}\, (M_C^2)_{JL\,}{}^{; \ua}
~+~ 
\bff_{IK\,}{}^{\ua}\, (M_C^2)_{JL\,; a} 
\Big) 
\overline{I}{}^a{}_{\ua}{}^{IJKL}(M^2,M_V^2,M_V^2)~. 
\labl{2LKh}
\end{eqnarray}
In this expression we introduced the \Kh\ curvature
$R^\ua{}_a{}^\ub{}_b$, and the third covariant derivatives $W_{;abc}$ of the
superpotential. Furthermore, we defined the full mass matrix for the
chiral multiplets $M = M_W + M_G$, with $(M_G^2)^a{}_b =
2\, (T_I \gf)^a\, (h\inv)^{IJ}\,  (\bgf T_J G)_b$ the mass of the
Goldstone bosons in the gauge \eqref{GFac}. 

Obvious in this talk we do not have time to discuss this computation
in detail, instead let us just mention one interesting fact. The first
term proportional to the curvature term is a result of supergraphs
with both the ``$8$'' and the ``$\ominus$'' topology, where the vertices
are the fourth and two third mixed derivatives of the \Kh\
potential. The covariant result involving the curvature is only after
realizing that in this case the ``$\ominus$'' supergraph turns into
scalar graph with the ``$8$'' topology.

\section{The Wess--Zumino model at two loops}

Because the two loop result \eqref{2LKh} above is obtained for general
gauged non--linear supersymmetric sigma models, it's expression is
rather complicated. We use the simplest supersymmetric theory in four
dimensions, the Wess--Zumino model, to illustrate our one and two
loop results. The \Kh\ potential of this model is trivial: $K ~=~ \bgf
\gf\,$, which means that the connection and curvature are all
zero. The superpotential is given by 
\begin{equation}  
W(\gf) ~=~ \frac 12\, m \, \gf^2 + \frac 1{3!}\, \gl\, \gf^3~,
\end{equation} 
with $m$ and $\gl$ complex parameters. The expressions for the
one and two loop \Kh\ potentials, \eqref{1LKh} and \eqref{2LKh} simplify to 
\begin{eqnarray}
K_{1L} &=& \dsp 
\frac 1{16 \gp^2}\,\frac 12\,  M_W^2\, 
\Big\{ 2 ~-~ \ln \frac {M_W^2}{\bgm^2} \Big\}~, 
\\
K_{2L} &=& \dsp 
\frac1{(16 \gp^2)^2}\, \frac 14\, M_W^2 |\gl|^2 \, 
\Big\{
-5 + 4 \, \ln \frac{M_W^2}{\bgm^2} - \ln^2 \frac{M_W^2}{\bgm^2} 
+ 12 \, \gk\big(\sfrac4{\sqrt 3} \big)
\Big\}~.
\labl{KHrenWZ}
\end{eqnarray}
with $M_W^2 ~=~ |m + \gl \, \gf|^2\,,$ and the function $\gk(x)$ can
be found in \cite{Nibbelink:2005wc,Ford:1992pn}.


\begin{thebibliography}{14}
\expandafter\ifx\csname natexlab\endcsname\relax\def\natexlab#1{#1}\fi
\providecommand{\enquote}[1]{``#1''}
\expandafter\ifx\csname url\endcsname\relax
  \def\url#1{\texttt{#1}}\fi
\expandafter\ifx\csname urlprefix\endcsname\relax\def\urlprefix{URL }\fi
\providecommand{\eprint}[2][]{\url{#2}}

\bibitem[Grisaru et~al.(1979)]{Grisaru:1979wc}
M.~T. Grisaru, W.~Siegel, and M.~Rocek, \emph{Nucl. Phys.} \textbf{B159}, 429
  (1979).

\bibitem[Novikov et~al.(1983)]{Novikov:1982px}
V.~A. Novikov, M.~A. Shifman, A.~I. Vainshtein, and V.~I. Zakharov, \emph{Nucl.
  Phys.} \textbf{B223}, 445 (1983).

\bibitem[Shifman and Vainshtein(1986)]{Shifman:1986zi}
M.~A. Shifman, and A.~I. Vainshtein, \emph{Nucl. Phys.} \textbf{B277}, 456
  (1986).

\bibitem[Seiberg(1993)]{Seiberg:1993vc}
N.~Seiberg, \emph{Phys. Lett.} \textbf{B318}, 469--475 (1993),
  \eprint{hep-ph/9309335}.

\bibitem[Intriligator~et al.(1994)]{Intriligator:1994jr}
K.~A. Intriligator~et al., \emph{Phys. Rev.} \textbf{D50}, 1092--1104 (1994),
  \eprint{hep-th/9403198}.

\bibitem[Groot~Nibbelink and Nyawelo(2006)]{Nibbelink:2005wc}
S.~Groot~Nibbelink, and T.~S. Nyawelo, \emph{JHEP} \textbf{01}, 034 (2006),
  \eprint{hep-th/0511004}.

\bibitem[Buchbinder et~al.(1994)]{Buchbinder:1994xq}
I.~L. Buchbinder, S.~M. Kuzenko, and A.~Y. Petrov, \emph{Phys. Lett.}
  \textbf{B321}, 372--377 (1994).

\bibitem[Buchbinder and Petrov(2000)]{Buchbinder:2000ve}
I.~L. Buchbinder, and A.~Y. Petrov, \emph{Phys. Atom. Nucl.} \textbf{63},
  1657--1670 (2000).

\bibitem[Ovrut and Wess(1982)]{Ovrut:1981wa}
B.~A. Ovrut, and J.~Wess, \emph{Phys. Rev.} \textbf{D25}, 409 (1982).

\bibitem[Grisaru et~al.(1996)]{Grisaru:1996ve}
M.~T. Grisaru, M.~Rocek, and R.~von Unge, \emph{Phys. Lett.} \textbf{B383},
  415--421 (1996), \eprint{hep-th/9605149}.

\bibitem[Pickering and West(1996)]{Pickering:1996gt}
A.~Pickering, and P.~C. West, \emph{Phys. Lett.} \textbf{B383}, 54--62 (1996),
  \eprint{hep-th/9604147}.

\bibitem[Buchbinder~et al.(2000)]{Buchbinder:1999jw}
I.~L. Buchbinder~et al., \emph{Nucl. Phys.} \textbf{B571}, 358--418 (2000),
  \eprint{hep-th/9906141}.

\bibitem[Brignole(2000)]{Brignole:2000kg}
A.~Brignole, \emph{Nucl. Phys.} \textbf{B579}, 101--116 (2000),
  \eprint{hep-th/0001121}.

\bibitem[Ford et~al.(1992)]{Ford:1992pn}
C.~Ford, I.~Jack, and D.~R.~T. Jones, \emph{Nucl. Phys.} \textbf{B387},
  373--390 (1992), \eprint{hep-ph/0111190}.

\end{thebibliography}
\end{document}